\documentclass{caosp302}
\usepackage{graphicx}

\articleNo{}
\pubyear{2008}
\volume{38}
\volnumber{3}
\firstpage{1}
\received{Oct 15, 2007}
\accepted{Oct 15, 2007}

\begin{document}

\hauthor{I.\,Iliev, J.\,Budaj}

\title{Am stars in binary systems}

\author{
        I.Kh.\,Iliev \inst{1}
      \and
        J.\,Budaj \inst{2} 
       }
\institute{
Institute of Astronomy, NAO Rozhen, Sofia, Bulgaria
         \and  
Astronomical Institute, Tatransk\'a Lomnica 05960, Slovakia,
budaj@ta3.sk
          }

\date{Oct 15, 2007}

\maketitle
\begin{abstract}
It is argued that apart from the well known dependence of 
the Am phenomenon on the mass, age (effective temperature, gravity) 
and rotation there is also a complex dependence on the orbital 
parameters in binary systems. 
This is why the generally accepted scenario in which
the Am star rotation plays a unique role needs to be revisited, 
the strong correlations between the rotation, 
orbital period and eccentricity need to be properly 
addressed and tidal effects taken into account.
Recent observations of Am stars in binary systems are reviewed.

\end{abstract}

\section{Introduction, general context and the standard scenario}

Chemically peculiar (CP) stars is a well know group of anomalous
stars on the upper main sequence (see e.g. Romanyuk 2007, Wahlgren 2004
for the most recent reviews).
It is generally accepted that the slow rotation of
these stars is the primordial cause of the Am and Ap peculiarity.
Slow rotation translates into the low rotationally induced
mixing and stable atmospheres void of disturbing turbulence.
Then radiatively driven microscopic diffusion
takes over and drives a vertical or horizontal stratification
of chemical elements and observed abundance anomalies.
At low effective temperatures CP phenomenon disappears due to
the deep convection zones, at hotter temperatures it is 
because of the strong stellar winds.
Large scale magnetic fields, if present, can strongly
affect the diffusion, observed abundance anomalies, and
differentiate between the magnetic Ap and nonmagnetic
Am stars.
Slow rotation is due to the magnetic breaking in single
magnetic stars or tidal synchronization in binary stars.

With this picture in mind one can explain many of the fundamental
observed properties of Am stars:
\begin{itemize}
\item
The slow rotation with the cutoff of about 100 km/s;
\item
abundance anomalies decreasing with the rotational velocity;
\item
high fraction of close binaries in Am stars;
\item
absence of Am binaries with very short orbital periods
($P_{orb}<1.2$ days, such systems rotate above the 100 km/s 
cutoff due to the spin-orbit synchronization and, thus,
the abundance anomalies disappear).
\end{itemize}

\section{The problems with the generally accepted scenario}

Nevertheless, there are inconsistencies and potential
problems with the above mentioned standard picture
in which rotation and diffusion play a unique role.
The picture invokes tidal breaking, synchronization,
circularization but ignores tidal effects on the stellar
magneto-hydrodynamics and mixing. It assumes that Am stars
rotate as single stars while most of them are close binaries.
Could it be that the Am anomalies depend primarily
on the orbital period or eccentricity which, in close binaries, 
strongly correlate with the rotation and this is why
we observe a correlation with the rotation?
Could it be that the correlation of Am
anomalies with the rotation and the rotation cutoff velocity
are not due to the rotationally
induced mixing but due to the tidal mixing which
might depend on the degree of the pseudo-synchronization
and thus on the rotation?
Could it be that the absence of short period Am binaries
is not due to their high rotation but due to the enhanced
mixing and flows in the strongly irradiated tidally distorted 
star with the surface approaching the Roche lobe?
To our knowledge nobody has ruled out such options so far.

There are more problems which is difficult to explain
within the standard picture.
Budaj (1996, 1999), Fe\v{n}ov\v{c}\'{\i}k et al. (2004),
and references therein studied the Am and Ap anomalies
as a function of rotation and orbital elements.
They found evidences that the properties of Am and Ap binaries 
depend on their orbital elements (some of them were questioned 
by Noels et al. 2004). It was suggested that the tidal effects 
in binary systems and a complex interplay between the binarity,
rotation and magnetism play crucial role in driving 
the CP phenomenon and are more `far-reaching'
than originally though.

There is a clear evidence of the enhanced Li abundances 
in the cool late type dwarfs and giants in close binaries 
indicating very different mixing in the envelopes of 
a tidally locked star than in a single star 
(Spite et al. 1994, Costa et al. 2002).

The tidal mechanism of slowing down the rotation of an Am star
is not well understood so far. There are two very different 
theories for the tidal synchronization and circularization.
The dynamical tide theory of Zahn (1977)
and the hydrodynamical mechanism of Tassoul \& Tassoul (1992).
The first mechamism can synchronize the spin of stars with
radiative envelopes for orbital periods up to a few days.
The latter is much more efficient and could reduce the stellar 
rotation up to orbital periods of 100 days.

Spin-orbit synchronization was typically observed in A V type 
binaries for orbital period of a few days. There is a tendency 
for pseudo-synchronization of Am binaries for orbital periods up 
to 30 days (Budaj 1996). However, recently Abt \& Boonyarak (2004) 
concluded that in BA IV or V binaries with periods as long as about 
500 days, the rotational velocities of the primaries are reduced 
relative to the primaries in wider binaries and single stars which 
is due to tidal effects.
Abt (2005) found out that binaries with B0–-F0 IV or V primaries 
with intermediate orbital periods (10-–100 days) lack highly 
eccentric orbits and that there is a tendency for circularization 
for periods up to about 1000 days.

Thus, there is no reason to believe that the magneto-hydrodynamics
in a member of the close binary is the same as in the single star.
Consequently, the observed properties of Am stars mentioned in 
the previous section cannot be considered as a proof of this
standard text book explanation of the Am phenomenon
and need to be revisited. Apart from the obvious
dependence of the Am phenomenon on the mass, age
(effective temperature, gravity) and rotation
the dependence on the orbital elements has to be studied also
and the strong correlations between the rotation, 
orbital period and eccentricity need to be properly 
addressed and taken into account.

\section{Am stars in binary systems}

A more detail study of Am stars in binary systems involving
the abundance analysis, determination of the orbital parameters, masses, 
radii, rotation, ages, and studies of synchonization and circularization
are thus very important. Recently, there have been a number of 
such studies.

Budaj \& Iliev (2003) studied three A-type binaries
and concluded that HD33254 is pronounced Am star,
HD198391 is extremely sharp-lined hot Am star,
and HD178449 is not Am star and that there is a faint
sharp-lined secondary spectrum. The original orbit
based on the photographic data turned out to be wrong.

Iliev et al. (2004) observed a few dozen SB1 Am binaries
hunting for the SB2 systems.
They detected the secondary spectra and estimated mass ratios
in HD434, HD861, HD108642, and HD216608.
In the last star they in fact observed three spectra and 
concluded that the previous orbit based on photographic data
was misinterpreted and wrongly assigned to the visual 
A component of the system.

Carquillat et al. (2004)
determined the orbit of 10 new Am binaries.

Lacy et al. (2004)
studied the detached, SB2, Am eclipsing binary star V885 Cyg
and determined masses, radii and effective temperatures.
The orbit is circular with the period 1.69 days, 
the observed rotational velocities are synchronous
with the orbital motion for both components.
The age of this system would seem to favor the hydrodynamic damping 
formalism of Tassoul \& Tassoul.

Mikul\'{a}\v{s}ek et al. (2004)
confirmed a mild Am-peculiarity of both components
of the double-lined spectroscopic eclipsing
binary HR 6611 with nearly circular orbit
and revealed a slightly asynchronous rotation
of the primary star.

North \& Debernardi (2004)
studied $e$ versus $P_{orb}$ dependence, and orbital period 
distribution of Am binaries and concluded that there are
two populations of Am stars.
Systems with $P_{orb}<30$ days owe their slow rotation to
tidal effects and systems with $P_{orb}>30$ days 
(or single stars) for which tides are not effective.

Fe\v{n}ov\v{c}\'{\i}k et al. (2004) studied Fe and Ca abundances 
in Am binaries as a function of the rotation, eccentricity, 
orbital period, and effective temperature.
They found that for orbital periods 10-200 days the abundance 
anomalies depend on the effective temperature, 
anti-correlate with the rotation, and correlate
with the eccentricity.

Kaye et al. (2004) found out that the SB2 binary, HD 221866,
is an Am (metallic-line A-type) star with 
the orbital period of 135 days.
The authors determined the basic physical 
and orbital parameters of both components.
They have similar masses, temperatures
and radii but the primary is the Am star, 
whereas the secondary appears to be a normal 
early F-type dwarf. 
However, it is the secondary which has lower vsini=14km/s 
than the primary vsini=19km/s.

Vuissoz \& Debernardi (2004) studied the distribution
of mass ratio, $q$, in Am binaries and found that it is 
centered on $q=0.56$.

Yushchenko et al. (2004) carried an abundance analysis
for both components of the SB2 star HD 153720.
Both components are Am stars and have similar
temperatures and projected rotation velocities.

Fr\'{e}mat et al. (2005)
analyzed a spectroscopic triple system DG Leo.
The inner binary consists of two Am components, 
at least one of which is not yet rotating
synchronously though the orbit is a circular one. 

Southworth et al. (2005) analysed eclipsing Am binary WW Aur
and determined precise masses, radii and temperatures
of both components and they are similar to each other.
Orbit has a period of about 2.5 days.
Synchronized rotation is in agreement with the theory of Tassoul
but not with that of Zahn.
The circular orbit of WW Aur is in conflict with the circularization
time-scales of both the Tassoul and the Zahn tidal theories. 

Iliev et al. (2006) studied 6 A-type binaries
and concluded that HD861, 29479, 108561 are typical Am
stars, HD20320 is a mild Am star, and HD18778
turned out not to be an Am star in spite of very low
vsini=27 km/s. On the contrary, HD96528 with vsini=85 km/s
is a mild Am star. The orbit of HD29479 based on
the photographic data is wrong.

Fekel et al. (2006) analyzed HR 1613 which is
a slowly rotating A type binary on nearly circular orbit,
orbital period of 8.1 days, vsini=11 km/s and equatorial 
rotational velocity of 30 km/s or less but without apparent 
Am anomalies. This is difficult to explain withing the standard
model but might be understood within the conclusion of 
Fe\v{n}ov\v{c}\'{\i}k et al. (2004) that the Am phenomenon 
is more pronounced in eccentric orbits.
On the contrary, $\theta And$ (vsini=93 km/s, Kocer et al. 2003)
seems to be an Am star.

Carquillat \& Prieur (2007) determined the orbital elements 
of 8 Am binaries and present statistical properties 
of the orbital parameters of the spectroscopic binaries
with an Am primary.

Zhao et al. (2007) carried out a comprehensive analysis
of $\lambda$ Vir and determined precise masses, orbit and
physical properties.
The masses and temperatures of both components are 
very similar but projected rotational velocities differ by 
a factor of 3-4. The orbit has a very small but nonzero 
eccentricity, orbital period of 206 days and nonsynchronous 
rotation of both components. The authors argue that 
this is in agreement with the tidal theory of Zahn (1977).

Fossati et al. (2007), Adelman \& Unsuree (2007),
Ryabchikova (2005) found a good agreement of
the Am abundance anomalies with the prediction of
the diffusion theory.

Burkhart et al. (2005) found no abundance trend
for Al, Si, S, and Fe during the Main Sequence evolution
of Am stars.

\section{Conclusion}

A more complex and detailed study of Am stars in binary systems
can shed more light on the fundamental questions
of the origin and nature of the CP stars as well as
on the interior hydrodynamics in binary systems.

\acknowledgements
II acknowledges the partial support from Bulgarian 
NSF under grant F-1403/2004. 
JB acknowledges the grant VEGA 2/6036/6.

\end{document}